# Shungite in view of neutron scattering

E.F().Sheka[1], I.Natkaniec[2,3], N.N.Rozhkova[4], K.Holderna-Natkaniec[3]

[1]Peoples' Friendship University of Russia, Moscow, Russia;
[2]Frank Laboratory of Neutron Physics, Joint Institute for Nuclear Research, Dubna, Russia;
[3]Adam Mickiewicz University, Department of Physics, Poznań, Poland;
[4]Institute of Geology Karelian Research Centre RAS, Petrozavodsk, Russia

**Abstract**. Recently suggested new concept of shungite carbon (Int. J. Smart Nano Mat. DOI: 10.1080/19475411) exhibits this raw material as a multi-level fractal structure of nanosize fragments of reduced graphene oxide (rGO). In view of the extraordinary importance of the rGO starting material for the current molecular graphene technology, the natural rGO deposits turns out to be quite challenging, making it highly necessary to prove the reliability of the proposed rGO concept of shungite. Once consistent with all the block of the available geological and physical-chemical data obtained during the last few decades, the concept nonetheless needs a direct confirmation in terms of the current graphene science. The first such acknowledgement has been received just recently when studying photoluminescence (PL) of shungite dispersions (arXiv:1308.2569v2). A close similarity of PL spectra of aqueous dispersion of shungite and those of synthetic graphene quantum dots of the rGO origin has been established. The current paper presents the next direct confirmation supplied with neutron scattering. Elastic neutron diffraction and inelastic neutron scattering have left no doubts concerning both graphene-like configuration and chemical composition of basic structural elements of shungite attributing the latter to rGO nanosize sheets with an average 11:1:3 (C:O:H) atomic content ratio. The experimental data are supplemented with quantum-chemical calculations that allowed suggesting a clear vision of the shungite structure at its first nanolevels.

## 1. Introduction

High-yield production of few-layer graphene flakes from graphite is important for the scalable synthesis and industrial application of graphene. Graphene-based sheets show promise for a variety of potential applications, and researchers in many scientific disciplines are interested in these materials. Although many ways of generating single atomic layer carbon sheets have been developed, chemical exfoliation of graphite powders to graphene oxide (GO) sheets followed by deoxygenation to form chemically modified reduced graphene oxide (rGO) has been so far the only promising route for bulk scale production. However, available technologies face a lot of problems among which there are low yield, the potential fire risk of GO and rGO when alkaline salt byproducts are not completely removed, a great tendency to aggregation, a large variety of chemical composition, and so forth (see the latest exhaustive reviews [1, 2] and references therein). In light of this, the existence of natural rGO is of utmost importance. As if anticipating the future need for the substance, the Nature has taken care of a particular carbon allotrope in the form of well-known shungite from deposits of carbon-rich rocks of Karelia (Russia) that strongly kept secret of its origin and rGO-based structure. Just recently has been suggested that shungite carbon has a multilevel fractal structure based on nanoscale rGO sheets [3] that are easy dispersible in water and other polar solvents [4, 5]. This suggestion has been the result of a careful analysis of physico-chemical properties of shungite widely studied to date as well as was initiated by knowledge accumulated by the current graphene science. Since then, two new direct justifications of the suggestion have been obtained. The first one is related to the study of photoluminescence (PL) of shungite aqueous and organic dispersions [6] that exhibits properties similar to those of synthetic graphene quantum dots of the rGO origin [7]. The second was obtained in the course of the neutron scattering study that is presented in the current paper. The study was initiated by two reasons. The first follows from the leading concept of the suggestion [3] that shungite was born in aqueous environment. Once porous due to its fractal structure, it provides favorable conditions for the water confining. Actually, thermal analysis and mass spectroscopy pointed the water presence up to a few wt% [8]. The other concerns a serious problem of detecting chemical composition of rGO, in general, and in shungite, particularly. Usually, the main attention is given to the determination of the remained oxygen contribution [2] while the hydrogen one was mainly attributed to the retained water. The current inelastic neutron scattering (INS) spectroscopy, which is the most hydrogen-sensitive technique, has allowed for the first time not only to detect the small-mass-content hydrogen component of the shungite carbon body but to suggest a general chemical formula for shungite rGO.

## 2. Experimental details

**Neutron experiment**. Neutron scattering study was performed at the high flux pulsed IBR-2 reactor of the Frank Laboratory of Neutron Physics of JINR by using the NERA spectrometer [9]. The investigated samples are illuminated by white neutron beam analyzed by time-of-flight method on the 110 m flight path from the IBR-2 moderator. The inverted-geometry spectrometer NERA allows simultaneous recording of both Neutron Powder Diffraction (NPD) and INS spectra. The latter are registered at final energy of scattered neutrons fixed by beryllium filters and crystal analyzers at $E_f = 4.65$ meV.

**Samples**. Three powdered shungite samples were subjected to the study. The first pristine shungite Sh1 presents the natural raw material with C ≥ 95 wt.% and grain size less than 40 μm from the Shun'ga deposits [4]. Shungite Sh2 was obtained when drying Sh1 at 110°C under ambient conditions for a week until the constant weight of the solid is reached. The evaporated water constitutes 4g per 100 g of the solid. The third shungite Sh3 presents a solid condensate of colloids of the shungite Sh1 aqueous dispersions and is produced in the course of the dispersion lengthy drying under ambient conditions until the constant weight is reached. Similarly to Sh1, additional heating of Sh3, sustained under ambient conditions for a long time, at temperatures above 100°C results in its releasing from the retained water constituting ~4% of the total mass. Raman scattering [3], high-resolution solid state $^{13}$C NMR [8], and Auger spectroscopy [5] show a deep similarity of shungites Sh1 and Sh3. Powdered spectral graphite was used to register both the reference NPD and INS spectrum from pure carbon material.

## 3. Neutron powder diffraction

Figure 1 presents a set of NPD plottings for the three shungite samples at 20 K. As seen in the figure, a general NPD pattern of all samples is identical and similar to that of spectral graphite while drastically differs from the latter by shape: all Gr(hkl) peaks are upshifted and considerably broadened pointing to irregular structure of the shungites. Similarly to the reference graphite spectrum, the main features of the shungite diffractograms are related to broad peaks located at ~3.5Å thus pointing to undoubted graphite-like pattern of structural packing, on one hand, and the packing close similarity for all the shungite samples, on the other. Figure 2 presents a detailed view on the main Gr(002) peaks. As seen in the figure, the narrow peak of graphite, the shape and width of which correspond to the resolution function of spectrometer and whose position determines $d_{002}$ interfacial distance between the neighboring graphite layers, is substituted with broad peaks whose characteristics are given in Table 1. To fix the position $A$ of the Gr(002) peaks maxima and to obtain the relevant FWHMs, $B$, the Gauss fitting procedure was applied. According to the fitting, the peak positions and FWHM of all shungite samples are identical within the limits of standard errors. The obtained data are listed in the table.

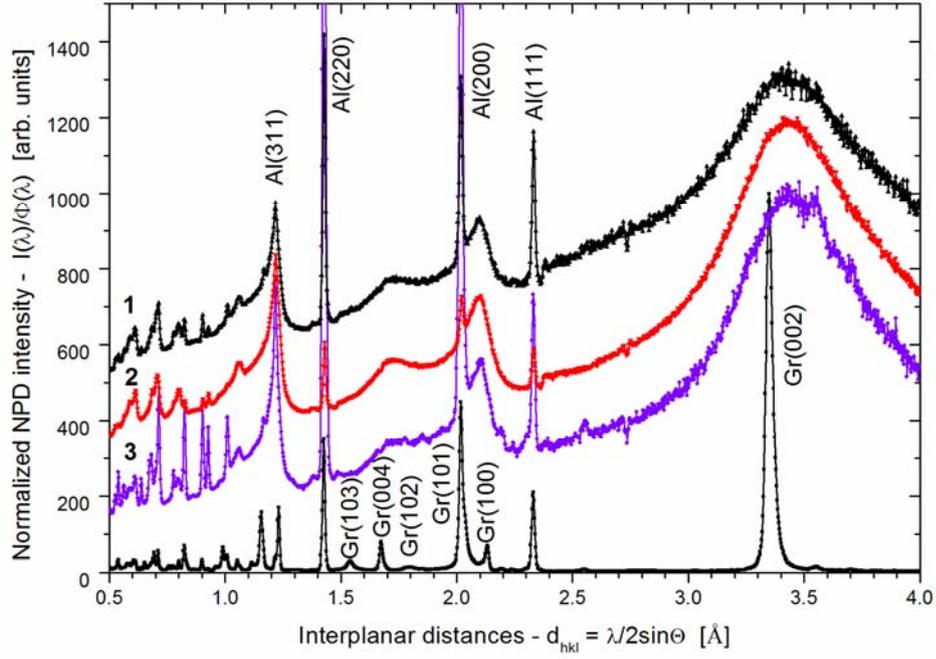

**Figure 1**. NPD of spectral graphite (Gr) and shungites Sh1 (1), Sh2 (2), and Sh3 (3) recorded at T=20K. Scattering angle $2\Theta = 117.4^\circ$. The data are normalized per neutron flux intensity $\Phi(\lambda)$ at each neutron wave length $\lambda$, next, intensity of both shungite and graphite peaks in the Gr(002) region are normalized to 1000 arbitrary units; Gr(hkl) and Al(hkl) denote characteristic diffraction peaks of spectral graphite and cryostat aluminum at different Miller indexes, respectively.

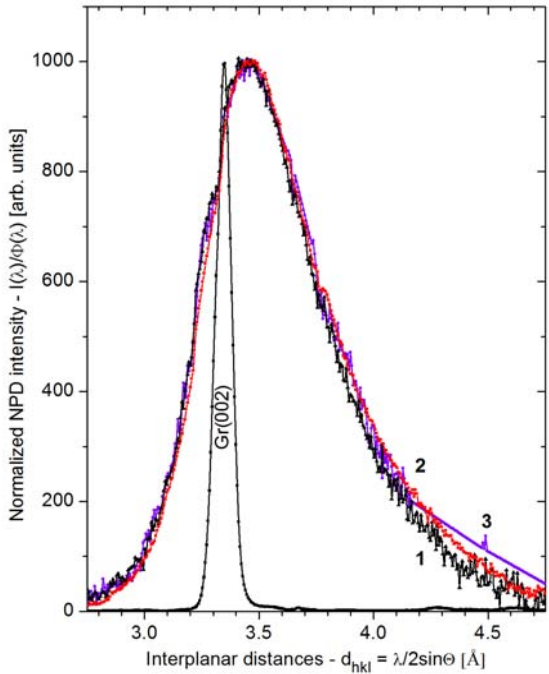

**Figure 2**. NDP fragments of spectral graphite (Gr) and shungites Sh1 (1), Sh2 (2), and Sh3 (3) in the region of the main diffraction peaks. T=20; scattering angle $2\Theta = 44.7^\circ$. The data are normalized per neutron flux intensity $\Phi(\lambda)$ at each wave length of incident neutrons $\lambda$; next after a linear background subtraction, the intensity maxima of all the peaks are normalized to 1000 arbitrary units.

Slight upshift of the peaks convincingly evidences a conservation of the graphite-like structure of all the shungite samples, while the peak wide broadening tells about a considerable space restriction. The latter is usually attributed to a narrowing of the coherent scattering regions (CSR) of a scatterer. According to widely used Scherrer's equation, the FWHM of a diffraction peak $B$ and the CSR length $L_{CSR}$ are inversely connected: $B = k\lambda / L_{CSR} \sin\Theta$, where $k$ is usually taken 0.94, $\lambda$ and $\Theta$ are the neutron wave length and scattering angle. When the diffraction study of a set of samples is performed under the same conditions, it is possible to take one of the samples as a reference one and to determine $L_{CSR}$ of the remaining samples addressing to that of the reference. In our study, $L_{CSR}^{ref}$ is attributed to crystalline graphite and constitutes ~20 $nm$ along both $c$ and $a$ directions [10]. Therefore, $L_{CSR}$ of the studied shungites can be determined as $L_{CSR} = (B_{ref}/B)(\lambda/\lambda_{ref})L_{CSR}^{ref}$. Substituting $\lambda/\lambda_{ref}$ by the ratio of the Gr(002) peak positions $A/A_{ref}$ and using $B$ values given in Table 1, we obtain $L_{CSR}$ = 1.3 $nm$. The data correlates well with those of 2.18 $nm$ and 2.30 $nm$ for Sh1 and Sh3, respectively, obtained by X-Ray diffraction [11]. Within the framework of shungite multilevel fractal structure [3], there might be two size limitations that correlate with the obtained $L_{CSR}$ values: these are 1) the linear dimension of individual rGO sheets (~1 $nm$) and 2) the thickness of the sheet graphite-like stacks that form the first level of the shungite structure. Summarizing NPD and X-Ray data, $L_{CSR}$ of ~1.5 - 2 $nm$ can be suggested. The latter implies ~5-6 layers of ~1 $nm$ rGO fragments in the stacks.. It should be noted that a multi-layer graphene-like packing is characteristic for rGO of any origin with $L_{CSR}$ along $c$ axis similar to the given above [1].

Table 1. Characteristics of Gr(002) peaks

| Samples | Peak position, $A$, Å | FWHM, $B$, Å |
|---|---|---|
| Graphite | 3.3501±0.0002 | 0.0341±0.0006 |
| Sh1 | | |
| Sh2 | 3.4513±0.0015 | 0.5408±0.0063 |
| Sh3 | | |

## 4. Inelastic neutron scattering

Figure 3 presents time-of-flight (TOF) INS spectra of the studied samples at T=20K. The spectra are summarized over 15 scattering angles, normalized per 10 hours exposition time. As seen in the figure, the intensity of the INS from graphite is at the level of the instrumental background and can be taken as the background one to be extracted from the shungite spectra for both experimental background and INS from carbon atoms to be excluded. The spectra clearly exhibit strong scattering from all the samples in contrast with that from graphite thus indicating that all of them are evidently hydrogen-enriched. At the same time, the spectra differ by both intensity and shape. Thus, if the spectra of Sh1 and Sh3 differ only in intensity, the spectra of Sh1 and Sh2 differ in both intensity and shape. The first finding points to that the hydrogen atom dynamics in Sh1 and Sh3 is rather identical while in Sh2 the later is quite different. Figure 4 presents the time-of-flight INS spectra of Sh1 and Sh2 at 20K. Since Sh2 was produced from Sh1 by lengthy heating, which was followed by removing water previously retained in Sh1, the difference spectrum 1-2 in the figure evidently presents the spectrum of the released water. Actually, well known characteristic features of the water INS spectrum are clearly observed in the spectrum of Sh1 while no traces of such structure are seen in the spectrum of Sh2. At the same time, the latter is quite intense, which undoubtedly points to the presence of hydrogen atoms incorporated in the carbon structure of shungite. The presence of the hydrogen atoms in the core of dried shungite has been directly observed for the first time. Therefore, the observed INS spectra of shungites are provided with incoherent inelastic neutron scattering (IINS) from hydrogens incorporated in the shungites structure and contained in the retained water.

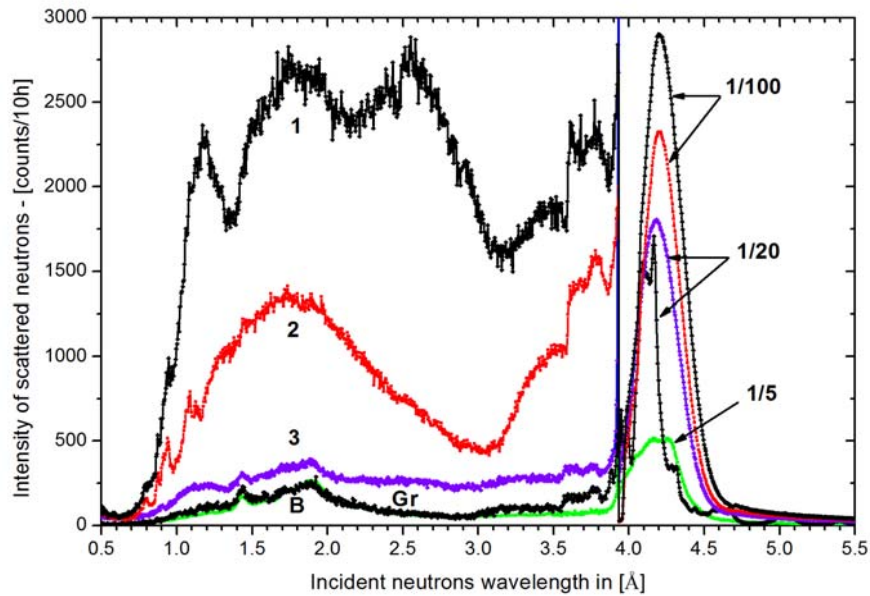

**Figure 3**. Time-of-flight INS spectra from shungites Sh1 (1=100g), Sh2 (2=96g), Sh (3=10g) and spectral graphite (Gr=10g). Curve B presents background from Al - cryostat and sample holder material. T=20K. The intensity of elastic peaks is 100-fold, 20-fold, and 5-fold reduced for Sh1 and Sh2, Sh3 and graphite, and background, respectively. Spectra are normalized per 10 hours exposition time at constant power of the IBR-2 equal 1.9 MW.

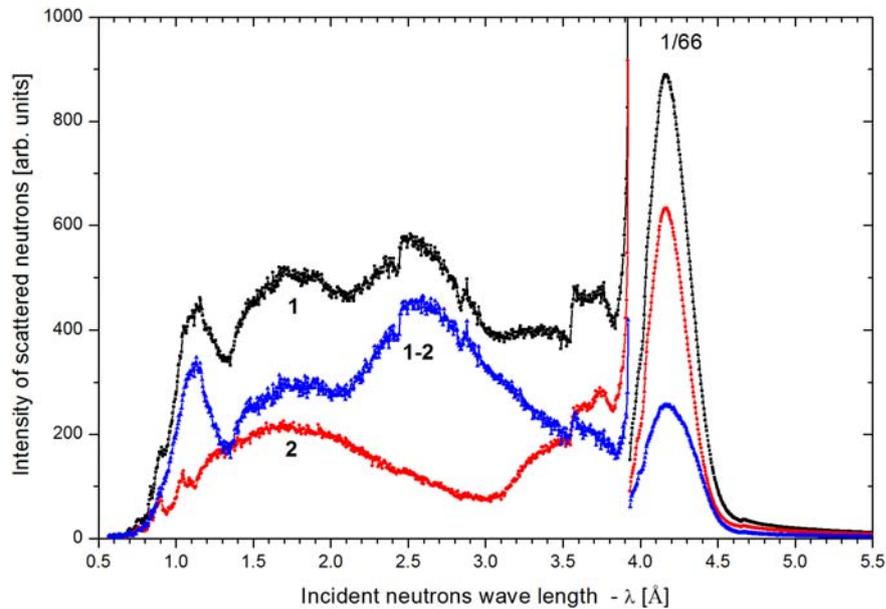

**Figure 4**. Time-of-flight INS spectra of shungites Sh1 (1) and Sh2 (2) at 20K after extraction of INS spectrum of graphite taken as background. Spectra are normalized per $10^6$ monitor counts of incident neutrons. Curve 1-2 presents the difference between spectra 1 and 2. The intensity of elastic peaks is 66-fold reduced.

Within the confines of commonly used incoherent one-phonon scattering approximation, the IINS spectra intensity is determined by the scattering cross-section (see for example Ref. 12)

$$\sigma_1^{inc}(E_i, E_f, \varphi, T) \approx \sqrt{\frac{E_f}{E_i}} \frac{\hbar |Q(E_i, E_f, \varphi)|^2}{\omega} \sum_n \frac{(b_n^{inc})^2}{M_n} \frac{\exp(-2W_n)}{1-\exp\left(-\frac{\hbar\omega}{k_B T}\right)} G(\omega). \qquad (1)$$

Here, $Q(E_i, E_f, \varphi)$ is the neutron momentum transfer; $\omega = (E_i - E_f)$ is the neutron energy transfer; $b_n^{inc}$ and $M_n$ are the incoherent scattering length and mass of the $n-th$ atom; $\exp(-2W_n)$ is the Debay-Waller factor; $G(\omega)$ presents the amplitude-weighted density of vibrational states (AWDVS) expressed as

$$G(\omega) = \sum_j \sum_n [A_j^n(\omega)]^2 \delta(\omega - \omega_j). \qquad (2)$$

Here, $A_j^n(\omega)$ is the amplitude of the $n-th$ atom displacement at the vibrational mode $\omega_j$.

Figure 5a presents one-phonon AWDVS $G(\omega)$ spectra obtained in the course of a standard treatment procedure [12] and related to the TOF spectra shown in Fig.4. The latter were not corrected on multi-phonon contribution that is quite small at low temperature. $G(\omega)$ spectra 1 and 3 are normalized per 100 g of mass. Changing in the fine structure of spectrum 3 above 500 $cm^{-1}$ may be connected with either poor statistic of experimental data due to small mass (10 g) of shungite Sh3 or reconstruction of its structure in the course of transformation of the pristine shungite Sh1 to Sh3 first through dispersion in water and then consolidation of a solid phase after the water evaporation. The differential spectrum 1-2 constituting the difference between spectra 1 and 2 as well as the spectrum of hexagonal ice is given in Fig. 5b.

Three main features follow from general overview of the spectra. The first concerns spectrum 1-2 in Fig. 5b that can be evidently considered as the spectrum $G^{wat}(\omega)$ of retained water. The spectrum presents the contribution of 4wt% water in spectrum of the pristine shungite $G^{Sh}(\omega)$ and is pretty similar to those well known for retained water in silica gels [13, 14], Gelsil glasses [15], oxygenated graphite [16], and various zeolites [17]. The second feature is related to the evident absence of the retained water crystallization so that the $G^{wat}(\omega)$ spectrum represents bound water [16], the molecules of which are connected with the solid shungite ground via hydrogen bonds and are located within the first adlayer. As shown in [18], this water may be supercooled without crystallization up to very low temperature. The third feature is related to the spectrum 2 in Fig. 5a that evidently exhibits hydrogen atoms in the shungite core, not connected with water. The relevant $G^{core}(\omega)$ spectrum differs drastically from the water one and is quite identical to the INS spectrum of a synthetic rGO [19] thus directly confirming the rGO nature of shungite. The spectrum is characterized by a considerable flattening up to 500 $cm^{-1}$ and reveals a pronounced structure in the region of 600-1200 $cm^{-1}$. The total intensity of $G^{core}(\omega)$ spectrum constitutes approximately 1/3 of the $G^{wat}(\omega)$ one. Taking into account that the latter is provided by 4wt% water, it is possible to evaluate 1.8wt% mass content of hydrogen in the shungite core. The data well correlate with those obtained by a thorough analysis of the chemical composition of synthetic rGO that reveals 1.5±0.5wt% hydrogen [20]. The two-component structure of hydrogen-enriched shungite takes place for shungite Sh3 as well as it follows from Fig. 5a.

As known, water molecules can be retained either on the surface of solid nanoparticles (see adsorbed water on aerosil [14]) and interfacial region of layered nanostructures [16] or in pores formed by the bodies [15,17]. The multi-layer graphene structure of shungites with interfacial spacing of ~3.45Å leaves no possibility of introducing 1.75Å-thick water molecules between the layers. The latter can be accommodated either on the outer surface of ~1.5-2 $nm$ shungite stacks or in pores made of the latter and their aggregates. Small angle neutron scattering (SANS) from shungite [21], which was possible precisely because of its hydrogen-enriched structure, exhibited pores of 2-10 *nm* and >20 *nm* in size thus pointing the place of the water accommodation. This explains a very deep similarity that is observed between the AWDVS spectra of the water retained in pores of silica gels [13, 14], Gelsil glass [13] and $G^{wat}(\omega)$ spectrum of shungites.

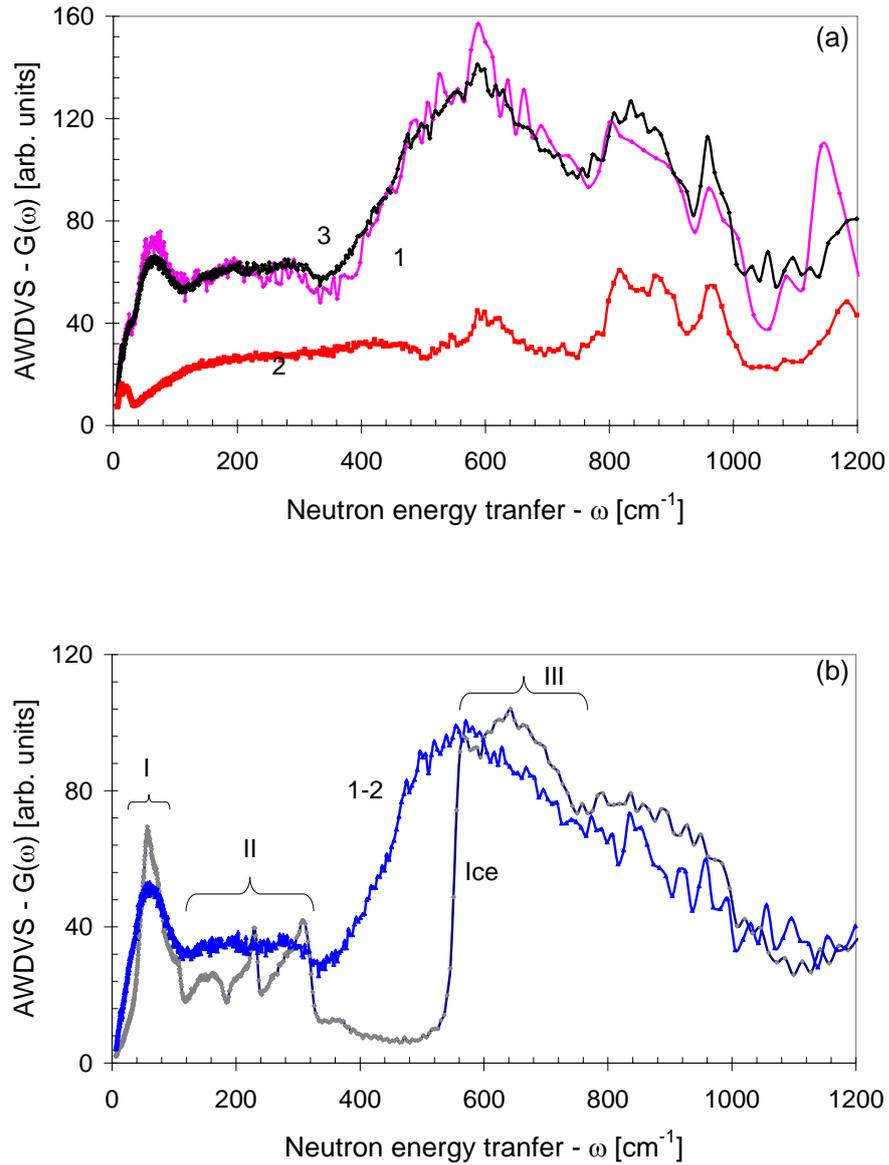

**Figure 5**. (a) One-phonon AWDVS spectra of shungites Sh1 (1), Sh2 (2), and Sh3 (3) at 20K.
(b) The difference spectrum (1-2) and spectrum of $I_h$ ice; T = 20K.

## 5. rGO sheet modeling

The two-mode hydrogen composition of shungite opens the way to clear the chemical covering of pore walls. According to the rGO concept of shungite structure [3], the pores are mainly covered with oxygen atoms of carbonyl units complemented by a small portion of hydroxyls [22]. However, as it has been noted even earlier [3], the relevant oxygen contribution is high which contradicts with real data indicating a small amount of oxygen [23]. Moreover, as follows from the $G^{core}(\omega)$ spectrum in Fig.5a, no indication of the presence of a band in the region of 90-100 $cm^{-1}$, which is characteristic for hydroxyls attached to solids [14], has been seen. In contrast, clearly vivid maxima at ~ 610 $cm^{-1}$, 820 $cm^{-1}$, 880 $cm^{-1}$, 960 $cm^{-1}$, and 1200 $cm^{-1}$, which fall in the region of the most characteristic non-planar deformational vibrations of the alkene C-H bonds, are observed. The hydrogen presence is supported by the hydrophobicity of rGOs that is often noted by chemists. The finding convincingly points to the presence of C-H bonds in the circumference of rGO nanosheets thus exhibiting a *post-factum* hydrogenation of the pristine rGO.

Available rGO models are tightly connected with those of pristine graphene oxide (GO) thus relating to the latter after removing all oxygen units from the sheet basal plane. Among a large variety of GO models, there is only one that is not just drawn basing on a chemical intuition but was obtained in the course of stepwise computational experiment subordinated to a particular algorithm [22]. Such a model is shown in Fig.6a. The (5, 5) GO molecule was computationally synthesized in the course of *per step* oxidation of the pristine (5, 5) nanographene (NGr) molecule (the latter is presented by a rectangular graphene fragment containing $n_a$=5 and $n_z$=5 benzenoid units along armchair and zigzag edges, respectively) in the presence of three oxidants, such as O, OH, and COOH. The equilibrium structure shown in Fig. 6b is related to the (5, 5) rGO molecule obtained due to removing the oxidants located at the basal plane of the (5, 5) GO molecule. This reduction mode is consistent with a remarkable difference in the *per step* coupling energies that accompany the attachment of the oxidants to either basal plane atoms (curves 1 and 2 in Fig.7c) or edge atoms in the circumference area (curve 3). Actually, the atoms located within the rose shading should be removed first. This evidently happens at the first stage of the real reduction and may present the final state of the reduction procedure when the latter is either short-time or not very efficient. This type of reduction might be attributed to a soft one. However, when the reduction occurs during long time or under action of strong reducing agents, it may concern oxidants located at the rGO sheet circumference due to a waving character of the *per step* coupling energy dependence with a large amplitude from -90 kcal/mol to -170 kcal/mol. Evidently, narrowing the interval occurs in the course of longtime and strong reduction, which is followed by removing oxidants. Thus, limiting the energy interval to 30 kcal/mol (removing oxidants covered by blue shading) results in remaining only 9 oxygen atoms (see Fig. 6d) in stead of 22 in the pristine (5, 5) rGO sheet shown in Fig.6b. Further stenosis of the interval in the limits of the cream shading in Fig.6c up to 20 kcal/mol leaves only 6 oxygen atoms (see Fig.6e). Consequently, the structures presented in Figs. 6b, 6d, and 6e might be attributed to rGOs obtained in the course of soft, medium, and hard reduction, respectively. The dependence of the chemical composition of final rGO products on the efficacy of the reduction process well explains C/O variation, a large scale of which is observed in practice depending on which namely chemicals are involved in the reduction procedure [2]. As for shungite, its formation during a long period of time allows for suggesting a hard type of the reduction occurred, which explains low oxygen content in the carbon-most-reach shungite deposits [23] and simultaneously provides stability of the chemical composition of the rGO basic elements.

Coming back to the rGO hydrogenation, we face question, whence comes the hydrogen. As discussed in [3], the hydrogenation of the pristine graphene lamellae loses to oxidation on all parameters in the course of the first stage of the graphenization of carbon sediments. Therefore, GO sheets similar to that shown in Fig.6a do not contain hydrogen among its framing atoms. The abundance of hot water around GO and rGO sheets, which accompanies the shungite derivation, suggests that water can provoke not only the GO reduction, but the hydrogenation of the formed rGO. Actually, the release of one of the edge carbon atoms of the rGO within the blue zone shown in Fig. 6c from oxygen makes the atom highly chemically active [24] and promotes the dissociation of a water molecule in the vicinity of this atom alongside with the neighboring oxygen. A possibility of such reaction in demonstrated in Fig.7. The water molecule, initially located at 1.10 Å apart from both the carbon and oxygen atoms in Fig. 7a, willingly dissociates (see Fig. 7b) while the formed hydroxyl remains in the vicinity of the newly formed C-H bond, once connected with both the bond hydrogen atom and neighboring oxygen via hydrogen bonds.

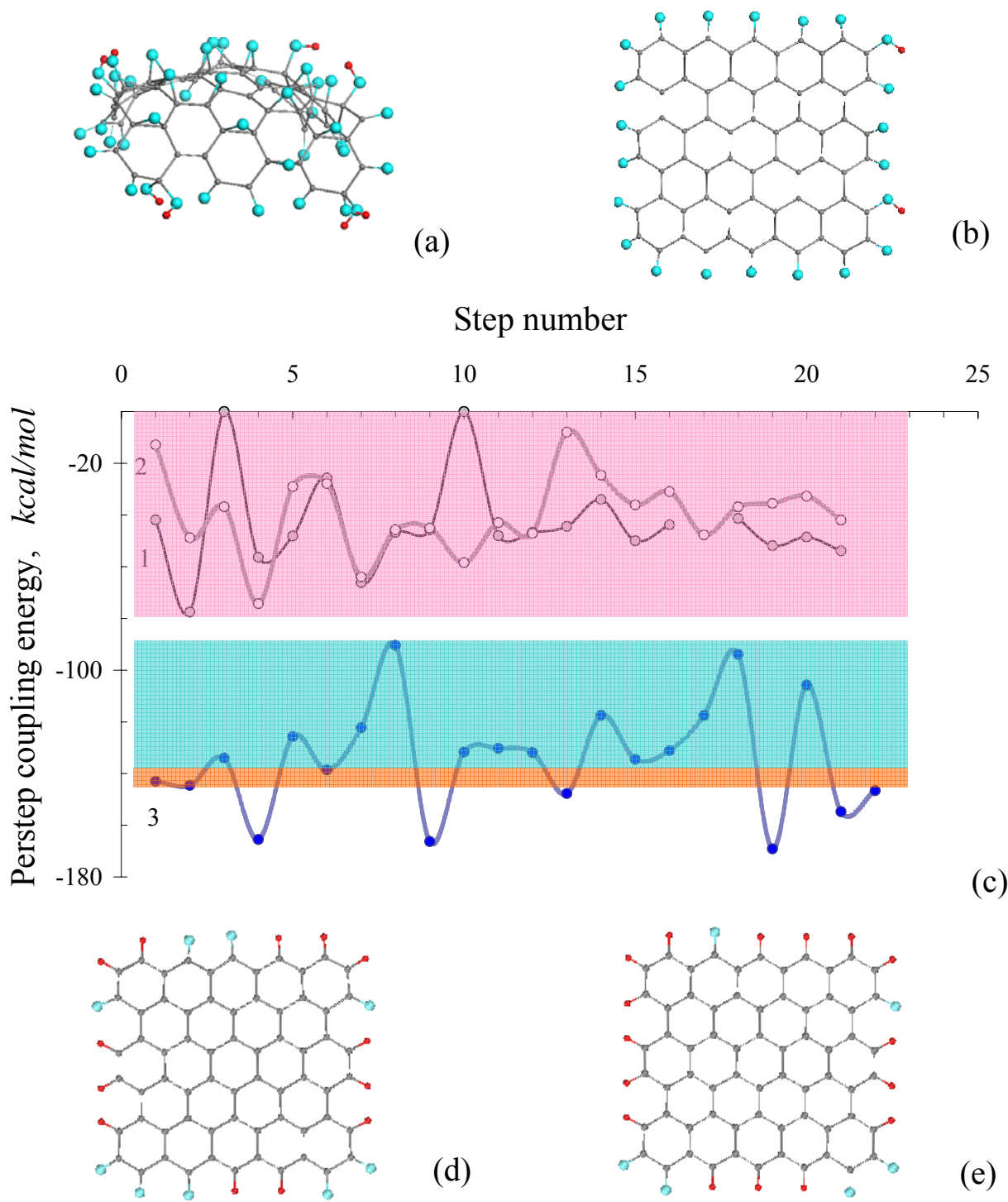

**Figure 6**. (a). Equilibrium structure of ~1 *nm* (5, 5) GO sheet corresponding to one-side oxidation of the pristine (5, 5) NGr molecule [21]. (b). Model rGO sheet corresponding to the first stage of the (5, 5) GO reduction affecting only the atoms in the basal plane of the GO sheet (soft reduction) [21]. (c). *per step* coupling energies related to the one-side oxygenation of the (5, 5) NGr molecule: O- and OH-attachments to the basal plane (curves 1 and 2, respectively) and the combination of O and OH attachments in the circumference (curve 3) [21]. (d) and (e) Model (5, 5) rGO sheets corresponding to a medium and hard reduction of the (5, 5) GO in the framework of the blue and blue-and-cream shaded zones in (c), respectively. Gray, blue and red balls present carbon, oxygen, and hydrogen atoms.

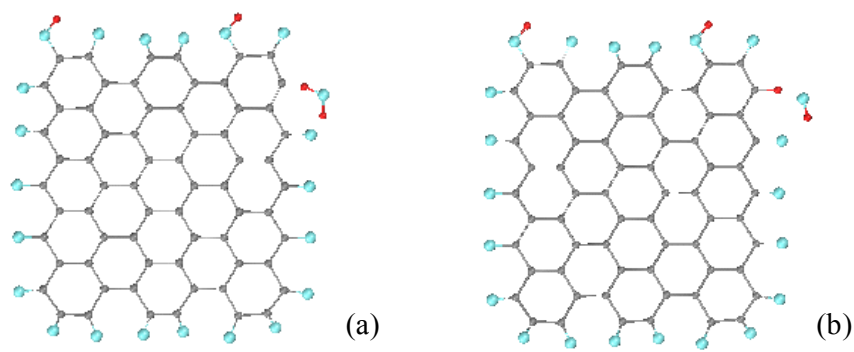

(a)  (b)

**Figure 7**. a. Starting configuration of the complex of one water molecule and the soft-reduced (5, 5) rGO sheet after removing one oxygen atom in the circumference. (b). Equilibrium structure of the (5, 5) rGO + water complex (UHF calculations). The atom colorings see in the caption to Fig. 6.

The energy gain of the reaction constitutes 25.94 *kcal/mol* which points to a high efficacy of the reaction. Evidently, the considered mechanism of the rGO hydrogenation, parallel or alongside with the simultaneous GO reduction, might not be the only one and will strongly depend on the reducing agents such as, for example, alcohols under critical regime [20]. However, hydrophobic character of produced rGOs, which is noted by many chemists, strongly evidences the reality of the pristine rGO hydrogenation.

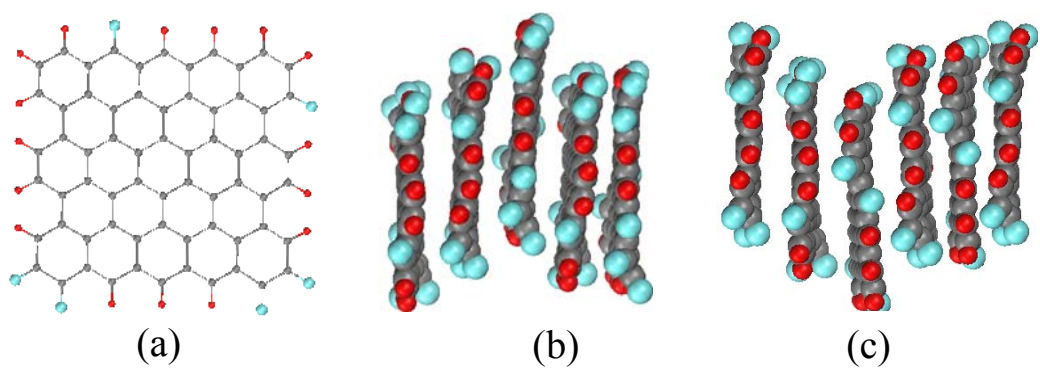

(a)  (b)  (c)

**Figure 8**. (a). Equilibrium structure of ~1 *nm* (5, 5) rGO sheet corresponding to a strong reduction of the pristine (5, 5) GO sheet [21]. (b) and (c). Arbitrary models of five- and six-layer rGO stacks with 3.5Å interlayer distance. Gray, blue and red balls present carbon, oxygen, and hydrogen atoms.

The chemical composition of the (5, 5) rGO sheet presented in Fig. 6e corresponds to the atomic and mass ratios listed in Table 2. It is consistent with the hydrogen content of Sh1 determined in the current study. As for experimental determination of the multicomponent content, the job is highly complicated and requires the usage of a complex of analytical techniques due to which the related information is rather scarce. Mainly, the analysis is restricted to the C/O ratio evaluation [2]. To our knowledge, there are only two more extended accounts, the data of which are presented in Table 2. As seen from the table, experimental data are in good agreement with that of suggested basing on the structure of the model (5, 5) rGO sheet presented in Fig. 6e. Relating to the chemical composition of the (5, 5) rGO involving 66 carbon atoms in total, 22 atoms of which are located in the circumference area, the chemical formula for the sheet can be presented as $C_{66}O_6H_{16}$, in particular, or ~$C_{11}O_1H_3$ ($C_{11}O_1H_{2.7}$), in general, that might be attributed to 11:1:3 (C:O:H) atom content ratio that should be attributed to hard reduced rGO sheets.

**Table 2**. Chemical composition of hard reduced rGO

| Ratio | C | O | H | Remarks |
|---|---|---|---|---|
| Atomic ratio, at% | 75 | 6.8 | 13.6 | Calc. |
| Mass ratio, wt% | 86.5 | 11.8 | 1.7 | Calc. |
| Mass ratio, wt% | 92.0±1.0 | 5.5±0.5 | 1.5±0.5 | Exp. [20] |
| Mass ratio, wt% | 85.7±1.0 | 9.59±0.5 | 1.06±0.5 | Exp. [25] |

## 6. Shungite pore modeling

Taking ~1 *nm* (5, 5) rGO sheet shown in Fig. 6e as the model basic element of the shungite structure as well as five- and six-layer stacks of the element, a model of shungite globules can be suggested. Presented in Fig. 8 is a planar vision of the globe with interglobe pores compatible with linear dimensions of rGO stacks.

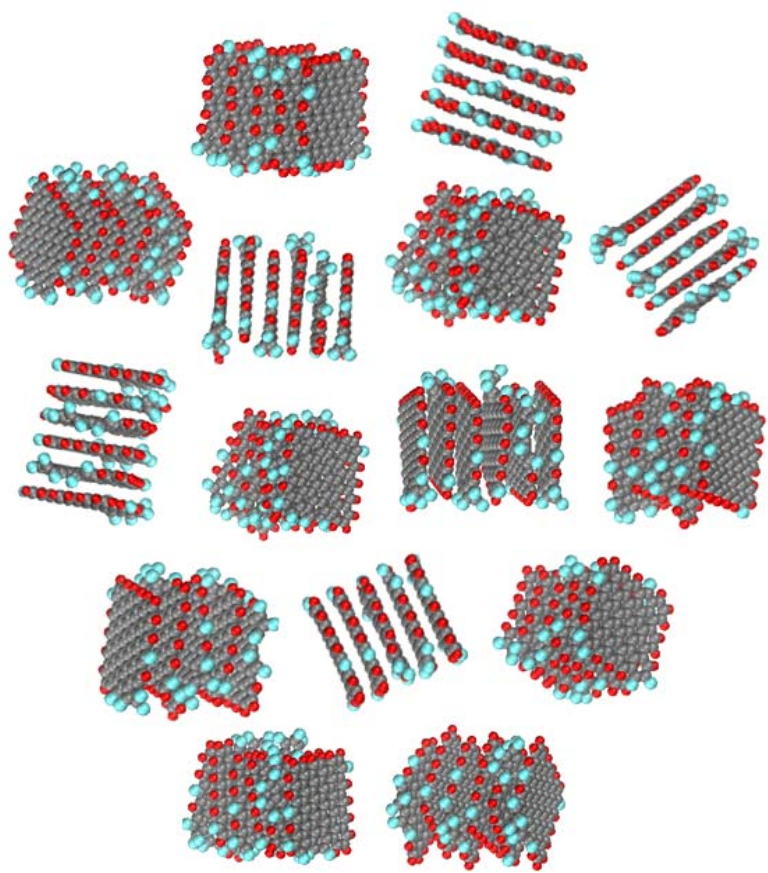

**Figure 9**. Planar presentation of shungite globule consisting of a set of five- and six-layer stacks of the (5, 5) rGO sheets voluntary located and oriented in space. Linear dimensions along the vertical and horizontal axes are of ~6 *nm*. The atom colorings see in the caption to Fig. 6.

As for the retained water, it is well known that in the low-frequency region (0-1000 cm$^{-1}$), the IINS spectrum of bulk water $G^{ice}(\omega)$ shows a hindered translational spectrum (I and II in Fig.5b)

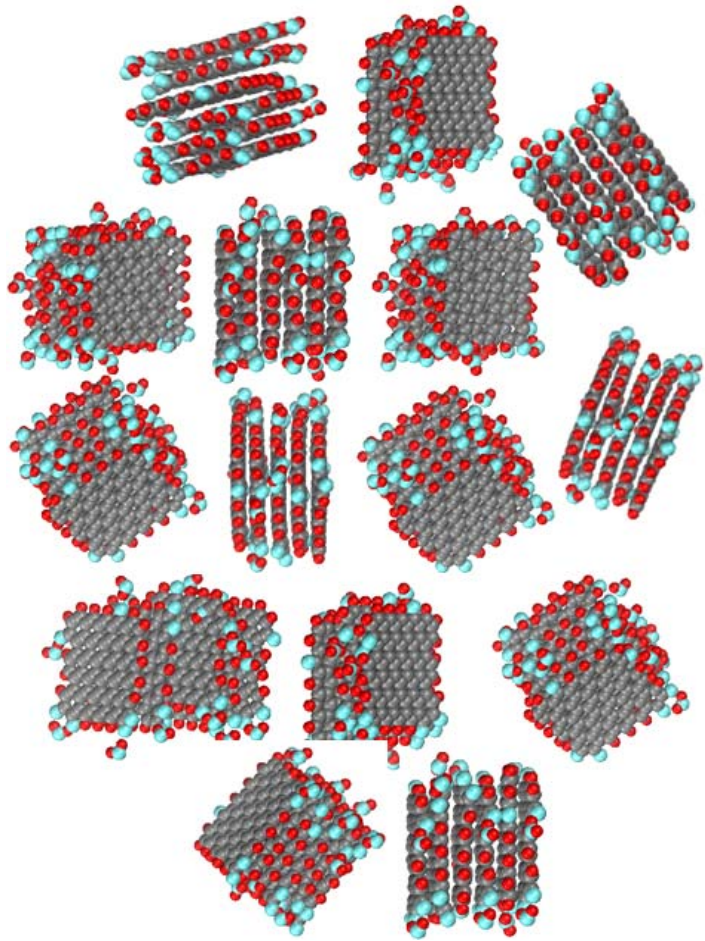

**Figure 10**. Planar presentation of shungite globules consisting of a set of five- and six-layer rGO stacks voluntary located and oriented in space and retaining two water molecules per one rGO sheet in average. Linear dimension along the vertical and horizontal are of ~6 *nm*. The atom colorings see in the caption to Fig. 6.

convoluted with the low-frequency Debye phonon-like acoustical contribution and a librational spectrum (III). Both the hindered translational and rotational (librations) modes are present in water because of intermolecular hydrogen bonds (HB) that are formed by each water molecule surrounded by other three ones. The configuration changes when water molecules can not move freely in space due to which it is quite natural to expect a strong spectral modification when passing from the bulk to retained water, which is clearly seen in Fig. 5b. Actually, the only ice mode positioned at ~56 *cm$^{-1}$* (HB bending) is retained in the $G^{wat}(\omega)$ spectrum while ~150 *cm$^{-1}$* (HB bending) as well as ~224 and ~296 *cm$^{-1}$* (HB stretchings) ice modes reveal clearly seen flattening and downshift. Analogous spectral modification takes place with respect to the ice librational modes forming a broad band in the region of 600-1200 *cm$^{-1}$*. The band is provided with water molecule rotations around three symmetry axes whose partial contribution determines the band shape. As shown by detail studies [15, 17], the modes conserve their dominant role in the IINS spectra of retained water, albeit are downshifted, when water molecules are coupled with the pore inner surface via hydrogen bonds. This very behavior is characteristic for the

$G^{wat}(\omega)$ spectrum in Fig. 5b. The three-ax partial contribution is sensitive to both chemical composition of the pore walls and the pore size [17]. Thus, the downshift of the red edge of the band from 550 $cm^{-1}$ to 320 $cm^{-1}$ when going from the $G^{ice}(\omega)$ spectrum to the $G^{wat}(\omega)$ one highlights the shungite pore size of a few *nm*, which is well consistent with SANS data [21].

A clear vision of the chemical composition of the basic structural elements of shungite as well as of their five- and six-layer stacking obtained on the course of the current study has allowed suggesting both a model structure of an individual shungite globule (Fig. 9) and a globule involving 4wt% water (Fig. 10), based on rGO sheet and its stacks shown in Fig. 8. In the latter case, the given mass content of water implies two water molecules per one (5, 5) rGO sheet in average. Actually, water can be distributed over the sheets irregularly, once concentrated on some sheets while leaving the other blank. This was taken into account when constructing the stack models in Fig. 10, just keeping two water molecules per one rGO sheet in average. As seen in the figure, the stacks with water molecules can be comfortably packed forming pores of a comparable size. The stacks surface, which forms the inner surface of the pores, is carpeted predominantly with hydrogens and much modestly with oxygens as components of framing carbonyl groups. Water molecules are predominantly accommodated near these very groups, once bound via HBs, thus limiting the monolayer coverage of the pores at much lower level than revealed by IINS study of graphene oxide [16].

## 7. Conclusion

The performed neutron scattering study proved to be extremely effective and put the last point in the study of the structure and chemical composition of shungite. The rGO nature of the basic structural element, its size and chemical composition as well as five-six-layer stacking have received a direct experimental confirmation fully supporting the general concept on the shungite structure suggested earlier [3]. Both the linear size of individual rGO sheets and the sheet stacks are responsible for the obtained $L_{CSR}$ values of 1.5-3 *nm*. The stacks of such dimension form globules of ~6 *nm* in size (the third level of structure) while the latter produce agglomerates of 20 *nm* and more (the forth level of structure) completing the fractal packing of shungite. The basic rGO fragments are hard reduced products of stable chemical composition described by 11:1:3 (C:O:H) atom content ratio due to which shungite deposits of Karelia present a natural pantry of highly important raw material for the modern graphene technologies.

**Acknowledgement**
The authors greatly appreciates financial support of RSF grants 14-08-91376 (E. Sh.) and 13-03-00422 (N.R.) as well as the Basic Research Program, RAS, Earth Sciences Section-5.